\documentclass{article}
\usepackage{spconf,amsmath,graphicx,hyperref}
\usepackage{booktabs} 
\usepackage{caption}
\usepackage{subcaption}
\usepackage{adjustbox}
\usepackage{standalone}
\usepackage{pgfplots}
\pgfplotsset{compat=1.18}
\usepgfplotslibrary{fillbetween}
\usepackage{amssymb}

\title{BAD: Taming the Bioacoustic Data Deluge with a Bat Activity Detector}

\name{Stefano Ciapponi$^{1,2}$, Santiago Martinez Balvanera$^3$, Andrea Cesaretti$^{2}$,  Elisabetta Farella$^1$, Kate E. Jones$^3$}
\address{Fondazione Bruno Kessler$^1$, University of Trento$^2$, University College London$^3$}

\begin{document}
\maketitle

\begin{abstract}
Passive Acoustic Monitoring of bats generates massive ultrasonic datasets ($>27\text{ GB/night}$ per node), straining edge storage and battery life. Legacy triggers fail against acoustic confusers, while deep models exceed microcontroller limits. We present a hardware-aware Bat Activity Detector (BAD) specifically designed to discriminate bat calls from hard biological and environmental confusers across variable sampling rates ($192\text{--}384\text{ kHz}$). Tailored for the Silicon Labs EFM32PG26 (MVP) in 8-bit integer precision, our model achieves 100\% hardware offload across all 14 layers ($17.2\text{ KB}$ Flash, $73.1\text{ KB}$ RAM). End-to-end preprocessing of ($74.00~\text{ms}$ for 76 frames) and inference ($30.00~\text{ms}$) of 100 ms clips at 192 khz require $104.00~\text{ms}$ per clip. On spatially out-of-domain recordings under a realistic low-prevalence regime ($r_{\text{pos}} = 0.05$), BAD achieves an AUC-ROC of 0.9748 and suppresses 99.4\% of non-target noise frames while retaining 65.3\% of bat calls—delivering a $>33\times$ precision gain over classical Goertzel baselines.
\end{abstract}


\section{Introduction}
\label{sec:intro}
Ultrasonic species monitoring, particularly in bat bioacoustics, faces a severe data deluge problem \cite{MEEIntegrating}. Passive Acoustic Monitoring (PAM) deployments rely on high sampling rates ($192\text{--}500\text{ kHz}$) to capture high-frequency echolocation calls. At $384\text{ kHz}$ (16-bit), a single node generates $\sim 2.76\text{ GB/h}$ ($>27\text{ GB/night}$), resulting in over $8\text{ TB}$ of raw audio for a standard recorder, month-long deployment. Crucially, the vast majority of this data contains no target biological activity or consists of non-target environmental noise.
While human speech processing relies on Voice Activity Detection (VAD) \cite{sohnvad, warden2018speech, dinkelTASLP} to filter idle audio, ultrasonic bioacoustics lacks an equivalent, low-power Bat Activity Detector (BAD) for edge microcontrollers (MCUs) \cite{saha2022tinyml}. Current deployments rely on classical triggers like amplitude thresholding or single-bin Goertzel filters \cite{goertzel1958algorithm}. Though computationally trivial, these methods fail in dynamic outdoor environments: amplitude thresholding triggers heavily on wind/rain, while Goertzel filters cannot distinguish frequency-modulated bat calls from ultrasonic confusers (e.g., orthopterans, small mammals, mechanical friction). This causes severe battery/SD card drain from recording noise, or missing soft calls entirely.
Deploying a tiny deep learning model on edge hardware addresses this by acting as an intelligent gatekeeper. Trained with hard negatives from non-target ultrasonic reference sets~\cite{middleton2020that}, our low-power BAD enables two primary deployment modes: \textit{(i) Low-Power Edge Logging,} suppressing non-target noise before high-energy SD card I/O writes occur; and \textit{(ii) Cascaded Pipelines,} acting as an efficient Stage-1 wake-up trigger for heavier downstream classifiers (e.g., BatDetect2~\cite{macaodha2022towards}) on high-tier hardware.

Although bioacoustic classification on microcontrollers is an active field \cite{VELASCOMONTERO2026103687, wrennet2026, benhamadi2026modeling, tinychirp2024}, and has recently gained community-wide momentum through dedicated TinyML bioacoustic benchmark challenges \cite{carmantini2025biodcase}, existing solutions rarely tackle the extreme sampling rates required for ultrasonic signals. To our knowledge, this is one of the first implementations to achieve high-throughput bioacoustic activity detection under these high-bandwidth conditions within constrained MCU memory. To ensure real-world applicability, our pipeline is designed around hardware specifications aligned with next-generation open-source bioacoustic platforms (e.g., upcoming AudioMoth revisions by Open Acoustic Devices\footnote{https://tinyurl.com/audiomothrevision}), specifically targeting MCU architectures with dedicated matrix acceleration. Deployed on a Silicon Labs EFM32PG26 MCU using Zephyr RTOS, our tiny CNN utilizes DMA double-buffering and zero-copy threading to ensure uninterrupted, drop-free execution.

\begin{figure}[t]
    \centering
    \includegraphics[width=1\linewidth]{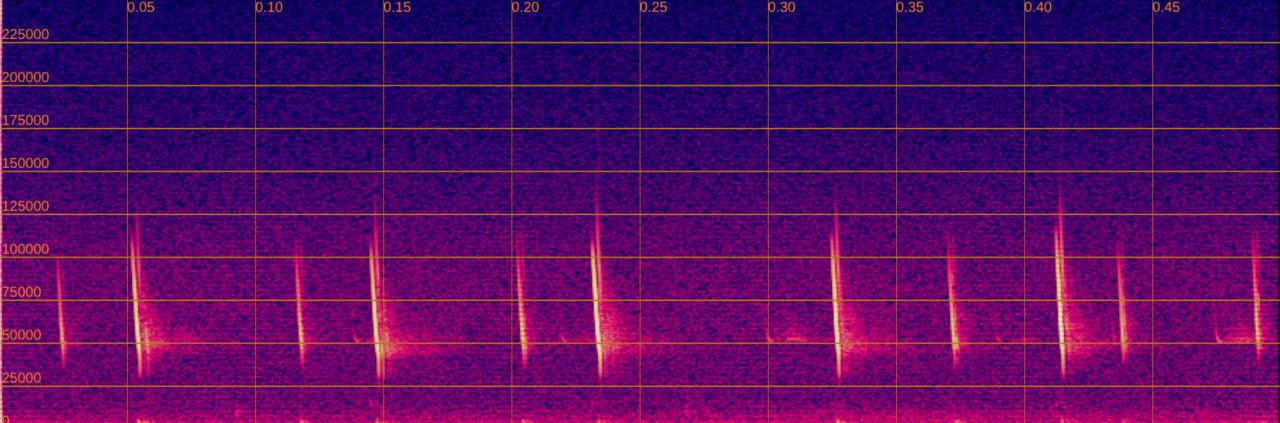}
    \caption{Spectrogram of a bat echolocation call sequence recorded at 256~kHz across a 0.5~s duration, illustrating rapid frequency-modulated (FM) sweeps within the 10--120~kHz band.\vspace{-0.5cm}}
    \label{fig:bat_call}
\end{figure}

In this work, we present a full system-algorithm co-design tailored for hardware-accelerated microcontrollers. Specifically, we evaluate the trade-offs of feature extraction parameters by benchmarking different temporal window lengths ($\tau \in \{50, 100, 200\}$~ms) and sampling rates against accuracy and RAM overhead. Furthermore, we provide detailed physical profiling of inference latency and active power/energy consumption to demonstrate field-readiness under continuous deployment constraints.

\section{Bat Echolocation Characteristics}
\label{sec:echolocation}
Bats emit short ultrasonic echolocation pulses—typically spanning 1–20~ms in duration with inter-pulse intervals ranging from 20~ms to over 100~ms during search phases~\cite{fenton1981echolocation, macaodha2022towards}, as showcased in figure \ref{fig:bat_call}. Individual call structures vary from steep frequency-modulated (FM) sweeps to constant-frequency (CF) components across 10–120~kHz. To reliably capture at least one complete vocalization cycle and its temporal context without incurring excessive hardware latency, bioacoustic processing relies on segmented windows. Our dynamic framing across $\tau \in \{50, 100, 200\}$~ms ensures that even transient FM chirps or sparse search-phase passes contain sufficient spectral energy for gating, while remaining compact enough for low-latency memory double-buffering on microcontrollers.

\section{Dataset Construction}
\label{sec:dataset}
We construct a balanced multi-source dataset of uniform $\tau = \{50, 100,200\}\text{ ms}$ clips to discriminate target bat vocalizations ($Class\,1$) from acoustic confusers ($Class\,0$). Positive samples ($Class\,1$) are extracted from BatDetect2 annotations~\cite{macaodha2022towards} centered at $t_{\text{center}} = (t_{\text{start}} + t_{\text{end}})/2$ with random temporal jitter $\delta \sim \mathcal{U}(-50\text{ ms}, 50\text{ ms})$. Negative samples ($Class\,0$) combine equal parts of: (i) non-overlapping background clips from BatDetect2, and (ii) hard biological and environmental negatives from the \textit{``Is That a Bat?''} dataset~\cite{middleton2020that} (owl chick begging, dormice, orthopterans, friction). Training batches ($N_{\text{train}} = 10{,}000$ clips/epoch) maintain a $1:1$ positive-to-negative ratio ($r_{\text{pos}} = 0.5$). To evaluate spatial generalization, we rely on two validation splits: \texttt{uk\_same}, using unseen audio from training sites, and \texttt{uk\_diff}, an out-of-domain split from geographically distinct recording sites with novel background noise profiles.

\begin{figure}[t]
    \centering
    \includegraphics[width=1\linewidth]{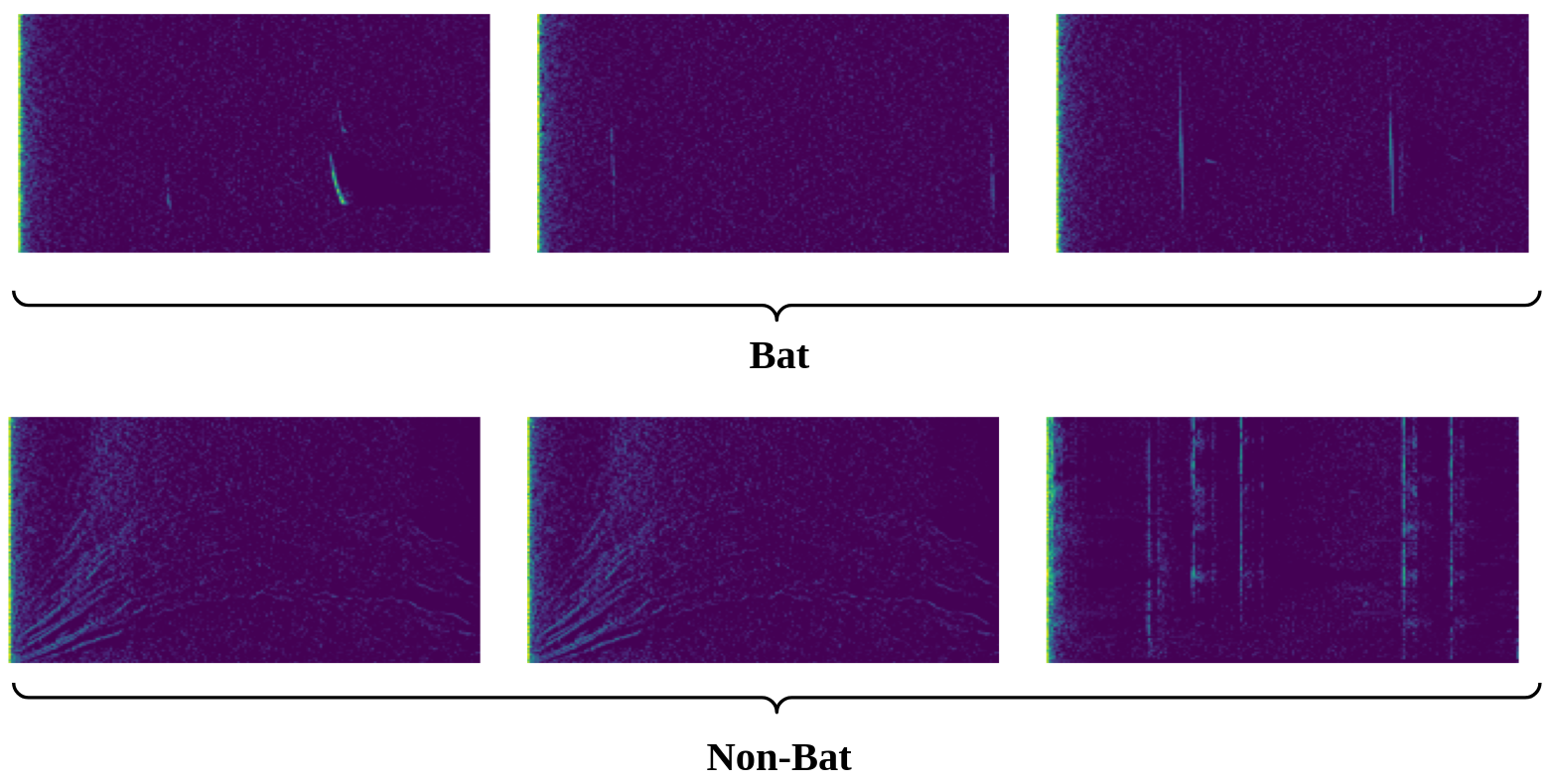}
    \caption{\textit{Bat} vs \textit{Non-bat} 200 ms clips resampled at 256Khz from Section~\ref{sec:dataset}, after preprocessing highlighted in Section~\ref{sec:preprocessing}.}
    \label{fig:bat_vs_non_bat}
    \vspace{-0.5cm}
\end{figure}

\section{Preprocessing \& Feature Extraction}
\label{sec:preprocessing}
To maintain adaptability across diverse hardware sensor setups, the audio preprocessing formulation is not bound to a single fixed sampling rate. Instead, the feature extraction pipeline dynamically adapts to input sampling frequencies $f_s \in \{192, 256, 384\}\text{ kHz}$ and variable clip durations $\tau \in \{50, 100, 200\}\text{ ms}$. 

Short-Time Fourier Transform (STFT) features, enabled by CMSIS-DSP, are computed using a fixed physical window duration $t_{\text{win}} = 2\text{ ms}$ ($N_{\text{fft}} = \lfloor f_s \cdot t_{\text{win}} \rfloor$, yielding 768 samples at $384\text{ kHz}$) with a $50\%$ overlap ($t_{\text{hop}} = 1\text{ ms}$) to preserve temporal resolution for rapid frequency-modulated (FM) sweeps \cite{sainath2015convolutional}. The resulting power spectrogram is bandpass cropped ($10\text{--}120\text{ kHz}$) and interpolated to $F = 64$ frequency bins. For a nominal input duration of $\tau = 200\text{ ms}$ at $384\text{ kHz}$, this results in $T = 134$ time frames.

To eliminate the heavy compute and memory overhead of Per-Channel Energy Normalization (PCEN) \cite{lostanlen2018per} on low-power microcontrollers, we implement a two-stage lightweight feature conditioning step:
\textit{(i) Background Tracking:} An Exponential Moving Average (EMA, $\alpha = 0.4$) filter updates a running background noise floor estimate with $O(1)$ memory state overhead.
\textit{(ii) Spectral Subtraction:} The background estimate is subtracted per spectral column to highlight short transient pulse events against stationary noise.
This pipeline yields an input tensor $\mathbf{X} \in \mathbb{R}^{1 \times 64 \times 134 \times 1}$ for model inference.

\begin{table*}[t]
\centering
\caption{Bat vs. Non-Bat Classification Validation Performance Across Sample Rates and Segment Lengths (Overall Best Metrics in \textbf{Bold}).}
\label{tab:bat_classification_results}
\resizebox{\linewidth}{!}{%
\begin{tabular}{ccccccccccc}
\toprule
\textbf{SR (kHz)} & \textbf{Segment (ms)} & \textbf{Epoch} & \textbf{Accuracy} & \textbf{Precision} & \textbf{Recall} & \textbf{Specificity} & \textbf{FPR} & \textbf{F1 Score} & \textbf{AUC-ROC} & \textbf{Avg Precision} \\
\midrule
192 & 50 & 9 & 0.7720 & 0.8908 & 0.6200 & 0.9240 & 0.0760 & 0.7311 & 0.8076 & 0.8587 \\
256 & 50 & 2 & 0.7410 & 0.9003 & 0.5420 & 0.9400 & 0.0600 & 0.6767 & 0.7484 & 0.8192 \\
384 & 50 & 5 & 0.7590 & 0.8453 & 0.6340 & 0.8840 & 0.1160 & 0.7246 & 0.7998 & 0.8538 \\
\midrule
192 & 100 & 5 & 0.9160 & 0.9298 & 0.9000 & 0.9320 & 0.0680 & 0.9146 & 0.9599 & 0.9684 \\
256 & 100 & 3 & 0.9030 & 0.9296 & 0.8720 & 0.9340 & 0.0660 & 0.8999 & 0.9495 & 0.9572 \\
384 & 100 & 16 & 0.9070 & 0.9111 & 0.9020 & 0.9120 & 0.0880 & 0.9065 & 0.9532 & 0.9626 \\
\midrule
192 & 200 & 6 & 0.9200 & 0.9167 & \textbf{0.9240} & 0.9160 & 0.0840 & 0.9203 & 0.9608 & 0.9594 \\
256 & 200 & 3 & \textbf{0.9300} & \textbf{0.9517} & 0.9060 & \textbf{0.9540} & \textbf{0.0460} & \textbf{0.9283} & \textbf{0.9710} & \textbf{0.9718} \\
384 & 200 & 14 & 0.9240 & 0.9435 & 0.9020 & 0.9460 & 0.0540 & 0.9223 & 0.9578 & 0.9593 \\
\bottomrule
\end{tabular}%
}
\end{table*}

\section{Model Architecture and HW Co-Design}
\label{sec:model_training}
The neural network architecture is explicitly tailored around the micro-architecture of the target MCU: the Silicon Labs EFM32PG26 featuring the Matrix Vector Processor (MVP) hardware accelerator. The MVP natively accelerates integer-quantized 8-bit (``int8'') operations—specifically standard 2D convolutions (\texttt{conv\_2d}), depthwise-separable convolutions \cite{chollet2017xception} (\texttt{depthwise\_conv\_2d}), 2D max pooling (\texttt{max\_pool\_2d}), spatial reductions (\texttt{mean}), and dense layers (\texttt{fully\_connected}).

To avoid falling back to costly host CPU software emulation, the network topology enforces full layer offloading onto the MVP hardware co-processor. The model backbone processes the input tensor $\mathbf{X} \in \mathbb{R}^{1 \times 64 \times 134 \times 1}$ through a stem stage followed by depthwise-separable stages: 
\textit{(i) Stem Layer:} A standard 2D Convolution ($3 \times 3$ kernel, 1 channel) followed by $2 \times 2$ Valid-padded Max Pooling downsamples the input tensor to shape $1 \times 32 \times 67 \times 1$; 
\textit{(ii) Depthwise-Separable Backbone:} Three cascaded depthwise-separable convolutional blocks expand channel capacity ($1 \rightarrow 16 \rightarrow 32 \rightarrow 32$) while progressively reducing spatial dimensions via $2 \times 2$ Max Pooling, where each block pairs a $1 \times 1$ pointwise Convolution (ReLU activation) with a $3 \times 3$ depthwise Convolution; 
\textit{(iii) Projection \& Spatial Reduction:} A final $1 \times 1$ pointwise Convolution expands the channel dimension to 64 ($1 \times 4 \times 8 \times 64$). Rather than using dynamic adaptive pooling layers which are unsupported on silicon accelerators, spatial dimensions are collapsed to $1 \times 1 \times 1 \times 64$ using an MVP-accelerated global \texttt{mean} reduction operator along the spatial axes; and 
\textit{(iv) Classification Head:} A final Dense (\texttt{fully\_connected}) layer projects the 64-dimensional embedding vector to 2-class ``int8'' logits representing \textit{Bat} vs.\ \textit{Non-Bat}.

Following standard microcontroller deployment practices \cite{banbury2020benchmarking}, the model is quantized to ``int8'' using Post-Training Quantization (PTQ) and trained for 20 epochs using Adam ($\eta = 10^{-3}$) with cross-entropy loss and label smoothing ($\epsilon = 0.1$):
\begin{equation}
    \mathcal{L} = -\sum_{c \in \{0, 1\}} y_c^{\text{smooth}} \log P(c \mid \mathbf{X}), \quad y_c^{\text{smooth}} = (1 - \epsilon)y_c + \frac{\epsilon}{2}
\end{equation}

\section{Experimental Evaluation \& Results}
\label{sec:evaluation}

\subsection{Feature Parameter Exploration}
\label{sec:feature_exploration}

To establish optimal feature extraction constraints prior to trigger comparison, we evaluate model performance across varying temporal segment lengths ($50$, $100$, and $200\text{~ms}$) and sampling rates ($192$, $256$, and $384\text{~kHz}$). The primary classification validation results are detailed in Table~\ref{tab:bat_classification_results}.

\textbf{Impact of Segment Length:} The duration of the input window serves as the primary driver of classification fidelity. Shorter windows ($50\text{~ms}$) consistently underperform across all sampling frequencies, reaching peak metric values of only $0.7720$ accuracy, $0.7311$ F1 score, and $0.8076$ AUC-ROC (at $192\text{~kHz}$). Extending the window duration to $100\text{~ms}$ yields a substantial gain, boosting AUC-ROC by over $15\%$ absolute ($0.9599$). Further expanding to $200\text{~ms}$ yields marginal additional gains, indicating that $50\text{~ms}$ segments lack sufficient temporal context, whereas $100\text{--}200\text{~ms}$ windows adequately capture full vocalization envelopes.

\textbf{Sample Rate Dynamics:}
Increasing $f_s$ from $192\text{~kHz}$ to $384\text{~kHz}$ does not monotonically improve accuracy. Across $100\text{~ms}$ and $200\text{~ms}$ windows, $256\text{~kHz}$ and $192\text{~kHz}$ configurations consistently match or outperform $384\text{~kHz}$ models while converging faster (Epoch $3$ vs.\ Epoch $14\text{--}16$). While $200\text{~ms}$ at $256\text{~kHz}$ achieves peak validation performance ($\text{AUC-ROC} = 0.9710$), the $100\text{~ms}$ window at $192\text{~kHz}$ offers an optimal edge trade-off, retaining an AUC-ROC of $0.9599$ and AP of $0.9684$ while halving the frame memory footprint.
\vspace{-0.3cm}
\subsection{Trigger Comparison Benchmarks}
\label{sec:trigger_benchmarks}

To evaluate trigger performance under the lowest memory overhead, we benchmarked all methods on $100\text{~ms}$ clips from the spatially out-of-domain \texttt{uk\_diff} split ($N=10{,}000$) across balanced ($r_{\text{pos}} = 0.5024$) and realistic low-prevalence ($r_{\text{pos}} = 0.0513$) regimes. Unseen recording sites force models to generalize across novel background noise profiles. Because false positives incur heavy battery and storage overhead on edge hardware, decision thresholds were optimized using the $F_{0.5}$ score ($F_\beta$ with $\beta=0.5$). Unlike $F_1$, $F_{0.5}$ weights Precision twice as heavily as Recall ($\beta^2 = 0.25$), explicitly penalizing energy-draining false wake-ups during high-noise periods.

As shown in Table~\ref{tab:benchmarks}, classical baselines struggle against outdoor acoustic confusers like rain, wind, and orthopterans. Amplitude thresholding (AMP) yields below-chance discrimination ($\text{ROC-AUC} \le 0.3915$) and excessive false alarms ($\text{FPR} > 76\%$), driving low-prevalence precision down to $5.70\%$. A Framed Goertzel Filterbank ($20\text{--}90\text{~kHz}$) improves overall ranking ($\text{ROC-AUC} = 0.6138$), but broad-band noise frequently triggers target bins, yielding a poor low-prevalence Average Precision (AP) of $0.0731$ and $F_{0.5}$ of $10.59\%$.


\begin{table}[b]
\vspace{-0.5cm}
\centering
\caption{Performance comparison on $100\text{~ms}$ clips under $F_{0.5}$-optimized decision thresholds ($N=10{,}000$, \texttt{uk\_diff} split).}
\label{tab:benchmarks}
\resizebox{\columnwidth}{!}{%
\small
\setlength{\tabcolsep}{3pt}
\renewcommand{\arraystretch}{1.1}
\begin{tabular}{l ccc ccc cc}
\toprule
\textbf{Method} & \textbf{AUC} & \textbf{AP} & \textbf{Prec} & \textbf{Sens} & \textbf{Spec} & \textbf{FPR} & \textbf{F1} & $\mathbf{F_{0.5}}$ \\
\midrule
\multicolumn{9}{l}{\textit{Balanced Prevalence ($r_{\text{pos}} = 0.5024$)}} \\
AMP        & 0.3915 & 0.4158 & 53.3\% & 86.1\% & 23.8\% & 76.2\% & 65.8\% & 57.7\% \\
Goertzel   & 0.6066 & 0.5845 & 57.1\% & 80.3\% & 39.0\% & 61.1\% & 66.7\% & 60.6\% \\
\textbf{BAD (Ours)} & \textbf{0.9732} & \textbf{0.9743} & \textbf{95.7\%} & \textbf{87.2\%} & \textbf{96.1\%} & \textbf{3.9\%} & \textbf{91.3\%} & \textbf{93.9\%} \\
\midrule
\multicolumn{9}{l}{\textit{Realistic Low Prevalence ($r_{\text{pos}} = 0.0513$)}} \\
AMP        & 0.3849 & 0.0385 &  5.7\% & 85.0\% & 24.0\% & 76.0\% & 10.7\% &  7.0\% \\
Goertzel   & 0.6138 & 0.0731 &  9.0\% & 36.1\% & 80.3\% & 19.7\% & 14.4\% & 10.6\% \\
\textbf{BAD (Ours)} & \textbf{0.9748} & \textbf{0.7785} & \textbf{85.7\%} & \textbf{65.3\%} & \textbf{99.4\%} & \textbf{0.6\%} & \textbf{74.1\%} & \textbf{80.7\%} \\
\bottomrule
\end{tabular}%
}
\end{table}

In contrast, our BAD model demonstrates robust spatial generalization ($\text{ROC-AUC} \ge 0.9732$). Under the $5\%$ prevalence regime, BAD rejects $99.41\%$ of noise frames ($\text{FPR} = 0.59\%$) while capturing $65.30\%$ of vocalizations. This achieves $85.68\%$ Precision and an $F_{0.5}$ of $80.65\%$, delivering a $>33\times$ precision gain over Goertzel filtering for long-term field deployment.

\vspace{-0.2cm}
\subsection{Hardware Profiling \& Energy Evaluation}
\label{sec:codesign}

The complete system is developed and deployed on the Silicon Labs EFM32PG26 microcontroller (ARM Cortex-M33 running at $80\text{~MHz}$ with MVP hardware accelerator). Continuous $192\text{~kHz}$ audio streaming is managed via Direct Memory Access (DMA) double-buffering under Zephyr RTOS.

To evaluate hardware offloading performance, the INT8 model was profiled using an Otii Arc Energy Profiler across 100 consecutive executions on physical hardware. Profiling confirms $100\%$ hardware offload efficiency across all 14 layers ($1.1\text{~M}$ MACs, $2.5\text{~M}$ total operations) with zero CPU fallback operations. The quantized model footprint requires $17.2\text{~KB}$ of Flash and $73.1\text{~KB}$ of runtime RAM (with $9.10\text{~KB}$ allocated for preprocessing buffers).

As detailed in Table~\ref{tab:performance_stats}, processing a nominal $100\text{~ms}$ audio clip ($19,200$ samples at $192\text{~kHz}$) generates exactly $76$ short-time feature frames given our STFT framing parameters (e.g., $512$-sample window with a $256$-sample hop size). Executing feature extraction across these $76$ frames requires $74.00\text{~ms}$ ($0.974\text{~ms/frame}$) alongside $30.00\text{~ms}$ ($0.395\text{~ms/frame}$) for neural network inference, achieving an end-to-end processing latency of $104.00\text{~ms}$. Physical power measurement during active MVP execution yields an average power consumption of $21.6\text{~mW}$ ($6.55\text{~mA}$ average current at $3.3\text{~V}$), corresponding to $180.0\text{~nWh}$ ($0.648\text{~mJ}$, $55\text{~nAh}$) per inference cycle.

\textbf{Real-Time Operation \& Scaling:} While the end-to-end processing latency for a $100\text{~ms}$ window ($\tau = 100\text{~ms}$) is $104.00\text{~ms}$ ($T_{\text{proc}} \approx \tau$), the minimal $4.00\text{~ms}$ overhead ($4\%$) is managed via DMA double-buffering under Zephyr RTOS, enabling near-real-time continuous operation with zero frame loss. Expanding the input duration to a $\tau = 200\text{~ms}$ audio window ($151$ frames) requires $141.03\text{~ms}$ for preprocessing and $59.60\text{~ms}$ for inference, bringing total latency to $200.63\text{~ms}$ while capturing peak classification fidelity ($\text{AUC-ROC} = 0.9608 $).

\begin{table}[t]
\centering
\caption{Stage Execution Times and Overall Processing Latency ($100\text{~ms}$ Audio Clips, $192\text{~kHz}$, $76\text{~Frames}$, $80\text{~MHz}$).}
\label{tab:performance_stats}
\scriptsize
\setlength{\tabcolsep}{3pt}
\renewcommand{\arraystretch}{1.1}
\begin{tabular}{l cccccc}
\toprule
\textbf{Metric} & \textbf{Windowing} & \textbf{Cropping} & \textbf{Resizing} & \textbf{SMS} & \textbf{EMA} & \textbf{FFT} \\
\midrule
\textbf{Per Frame ($\mu\text{s}$)} & $52.63$ & $65.79$ & $39.47$ & $39.47$ & $184.21$ & $592.11$ \\
\textbf{76 Frames ($\text{ms}$)} & $4.00$ & $5.00$ & $3.00$ & $3.00$ & $14.00$ & $45.00$ \\
\midrule
\midrule
\textbf{Metric} & \multicolumn{2}{c}{\textbf{Total Prep ($\text{ms}$)}} & \multicolumn{2}{c}{\textbf{Inference ($\text{ms}$)}} & \multicolumn{2}{c}{\textbf{Total Latency ($\text{ms}$)}} \\
\midrule
\textbf{Duration} & \multicolumn{2}{c}{$74.00$} & \multicolumn{2}{c}{$30.00$} & \multicolumn{2}{c}{$104.00$} \\
\bottomrule
\end{tabular}
\end{table}

\textbf{Deployment Considerations \& Duty-Cycling:}
While this work evaluates continuous frame-by-frame inference, the system architecture natively supports an event-driven duty-cycled workflow. In practical field deployments, the neural trigger can serve as a gating mechanism: upon detecting a target vocalization, the system can temporarily suspend inference and stream raw audio directly to storage for a fixed window (e.g., $3\text{--}5\text{~s}$) to capture passing bat sequences. By eliminating redundant neural processing during active passes and reverting to low-power monitoring thereafter, this strategy—planned for future field trials—will further extend operational battery life. At an energy cost of just $180.0\text{~nWh}$ ($0.648\text{~mJ}$) per inference, the neural trigger enables over $5.4 \times 10^7$ active evaluation cycles on a single standard $3.0\text{~Ah}$ lithium battery, demonstrating that the accuracy gains of deep edge-gating do not compromise field deployment longevity.

\section{Conclusion and Future Works}
\label{sec:conclusion}
In this work, we presented a system-algorithm co-design of a hardware-accelerated Bat Activity Detector (BAD) to address the bioacoustic data deluge in passive acoustic monitoring. By replacing fragile thresholding filters with a lightweight INT8 CNN, BAD gates ultrasonic streams directly at the edge. The model achieves $100\%$ hardware offloading across all 14 layers onto the Silicon Labs EFM32PG26 MVP accelerator, using $17.2\text{~KB}$ Flash and $73.1\text{~KB}$ RAM. On out-of-domain recordings under realistic low prevalence ($r_{\text{pos}} = 0.05$), BAD rejects $99.4\%$ of non-target noise frames ($\text{AUC-ROC} = 0.9748$), delivering a $>33\times$ precision gain over Goertzel baselines.

Hardware profiling confirms deployability: a $100\text{~ms}$ window ($76$ frames) operates in a low-latency near-real-time pipeline ($74.00\text{~ms}$ preprocessing + $30.00\text{~ms}$ inference = $104.00\text{~ms}$ total latency), maintaining zero-copy execution under DMA double-buffering. Requiring just $180.0\text{~nWh}$ ($0.648\text{~mJ}$) per inference, the trigger enables over $5.4 \times 10^7$ active cycles on a $3.0\text{~Ah}$ lithium battery.

Future work will focus on in-situ field testing across natural ecosystems once next-generation open-source hardware prototypes become available. These trials will evaluate long-term battery performance, seasonal environmental robustness, and adaptive duty-cycling schemes to further minimize power consumption.
\newpage
\section*{Acknowledgment}
The authors would like to thank Alex Rogers and the Open Acoustic Devices team for valuable technical insights and for sharing hardware specifications regarding next-generation AudioMoth platform targets.

\bibliographystyle{IEEEbib}
\bibliography{refs}

\end{document}